\documentclass{ws-procs9x6}
\usepackage{epsfig}
\usepackage{cite}
\begin{document}

\title{HELAC - A Monte Carlo generator for multi-jet processes }

\author{C. G. Papadopoulos}
\address{Institute of Nuclear Physics, NCSR ``Demokritos''\\
15-310 Athens, Greece\\
 E-mail: Costas.Papadopoulos@cern.ch}

\author{M. Worek 
\footnote{\uppercase{P}resented at the  \uppercase{XIV} 
\uppercase{I}nternational \uppercase{W}orkshop
on \uppercase{D}eep \uppercase{I}nelastic \uppercase{S}cattering 
\uppercase{(DIS2006)}, \uppercase{T}sukuba, \uppercase{J}apan,
20 $-$ 24 \uppercase{A}pril 2006. }}

\address{Institute of Nuclear Physics, Polish Academy of Sciences\\
Radzikowskiego 152, 
31-342 Krakow, Poland
\\E-mail: Malgorzata.Worek@desy.de}

\maketitle

\abstracts{  The status of  the multi-purpose event  generator
{\tt HELAC} is briefly presented. The aim of this tool is the  full
simulation of events within  the Standard Model (SM) at current and
future high energy experiments,  in particular the LHC. Some results
related to the production of  multi-jet final states at the LHC are
also discussed.  }

\section{Introduction}
The possibility of identifying new physics relies on predictions
of multi jet final states at current and future collider
experiments like TeVatron or LHC.  Signals for many models beyond the SM
involve a large number of jets resulting from decay chains of
particles with high masses.  Their Monte Carlo (MC) simulation,
both for signal and background is of crucial importance for
the success of experiments. One of the  ways to simulate multi jet
events is to use exact matrix elements at some  given
order of perturbation theory in the strong coupling constant $\alpha_s$.
The matrix
elements for fully exclusive final states for many jets which  we have
at our disposal in most cases are at leading order. 
The advantage of this
approach consists of having exact results with all interference
effects taken into account  properly. The main disadvantage is that
more and more diagrams have to be considered when the number of external
particles increases. Roughly speaking the number of Feynman diagrams
grows asymptotically factorially with the number of external particles.

In this report we will focus on tree level calculations. However,
recursive methods based on Dyson-Schwinger recursive equations are
used during calculations, resulting in the computational cost of $3^n$
comparing to  $n!$ in the traditional Feynman diagrams approach.

\section{Short description of the program}
{\tt HELAC} is a {\tt FORTRAN} code which  automatically constructs
helicity amplitudes using recursion techniques based on
Dyson-Schwinger equations. It has been
 introduced in  Ref.\cite{Draggiotis:1998gr} and recently revived in
Ref.\cite{Kanaki:2000ey,Kanaki:2000ms,Draggiotis:2002hm,
Papadopoulos:2005vg,Papadopoulos:2005jv,Papadopoulos:2005ky}.  The
initial and final state particles are given by the user. Subsequently, the
program calculates the corresponding sub-amplitudes  which 
contribute to the process under consideration 
and evaluates the total amplitude, without referring
to individual Feynman diagrams.   Apart from that, the
summation over helicity and colour configurations, usually
time consuming, is performed by  MC techniques,  see
Ref.\cite{Papadopoulos:2005ky} for details. This results  in the
computational cost of $3^n$, where $n$ is a number of the external
particles,  compared to  $n!$ in the traditional Feynman diagrams
approach. The program is able to calculate the matrix element for any
tree level  SM process.  The phase space sampling
subroutines,  see Ref.\cite{Papadopoulos:2005ky} for details, are also 
implemented to generate total rates as well as partonic  events.   
The program also includes an
acceptance-rejection part to unweight those events. The code is able
to generate a final state configuration made of  hard  quarks, gluons
and other non-coloured particles. The final state  is thus not
directly comparable to what is observed in the experiment. An event
generator that aspires to give a realistic description of collision
processes must include a way to compute/estimate the   effects of
higher order corrections in perturbation theory and describe
hadronisation effects. For parton  showering and
translation of partons into hadrons the code is interfaced   to
the latest version of  {\tt PYTHIA 6.4} \cite{Sjostrand:2006za} in a
standard way for {\tt FORTRAN} based event generators  by the Les
Houches Accord (LHA) event record \cite{Boos:2001cv}.

\section{Basic results}
\begin{figure}[h!]
\begin{center}
\epsfig{file=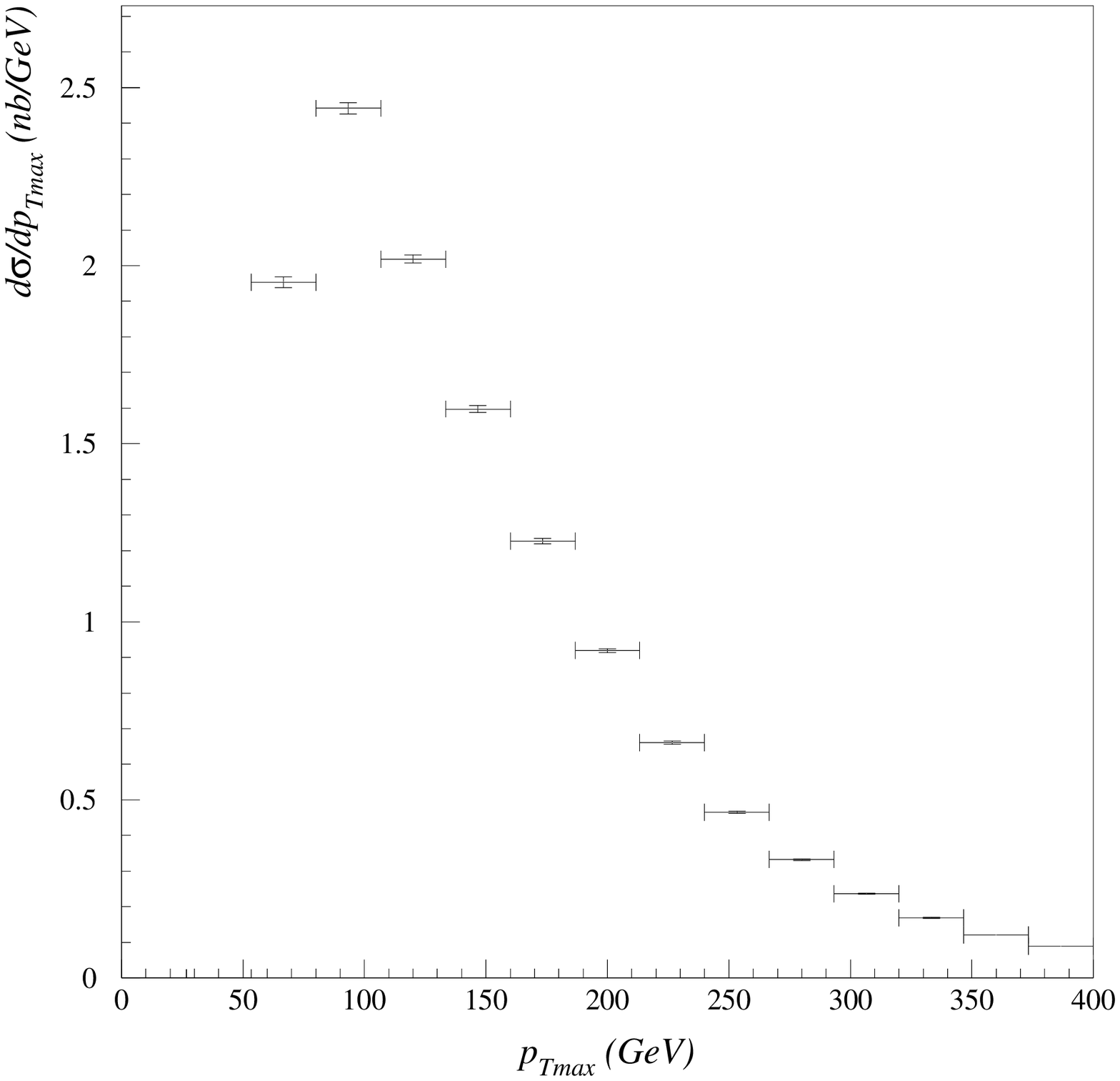,width=55mm,height=35mm}
\epsfig{file=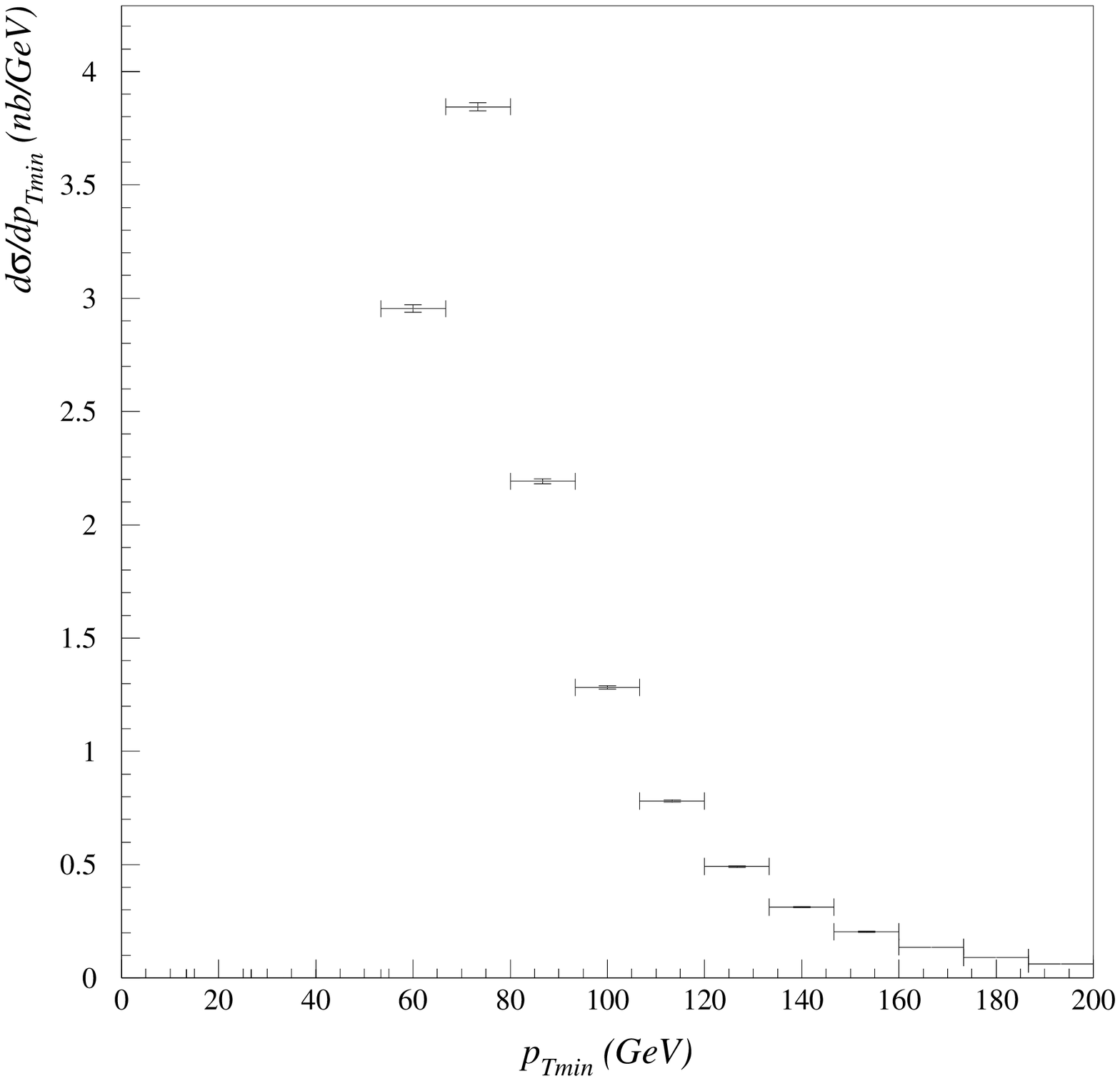,width=55mm,height=35mm}
\end{center}
\caption{
Transverse momentum distribution of the most (left panel)
and the least (right panel) energetic gluon in the $gg\rightarrow 4g$ process
with the MC summation over colour.}
\label{pt_6g}
\end{figure}

\begin{figure}[h!]
\begin{center}
\epsfig{file=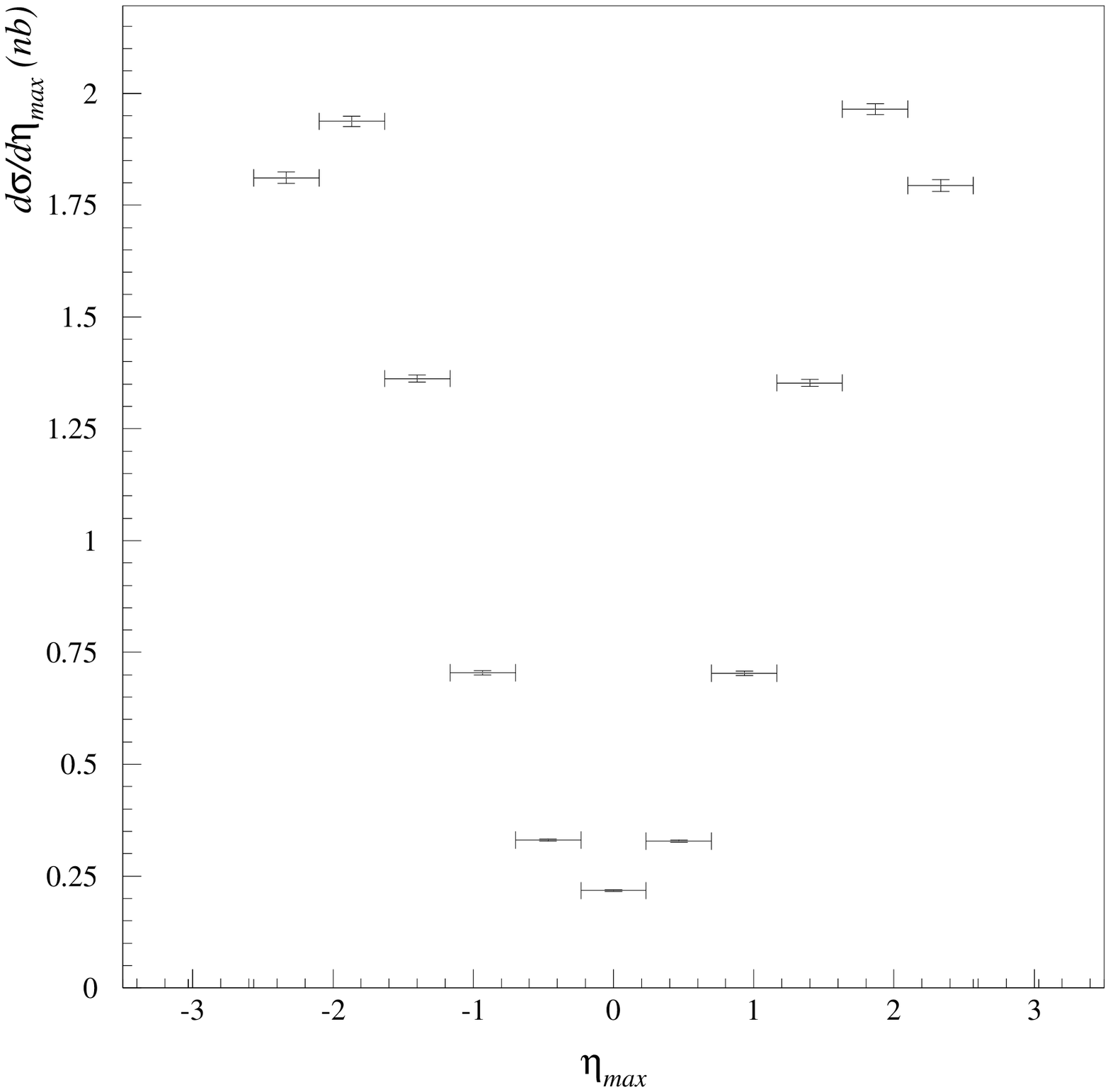,width=55mm,height=35mm}
\epsfig{file=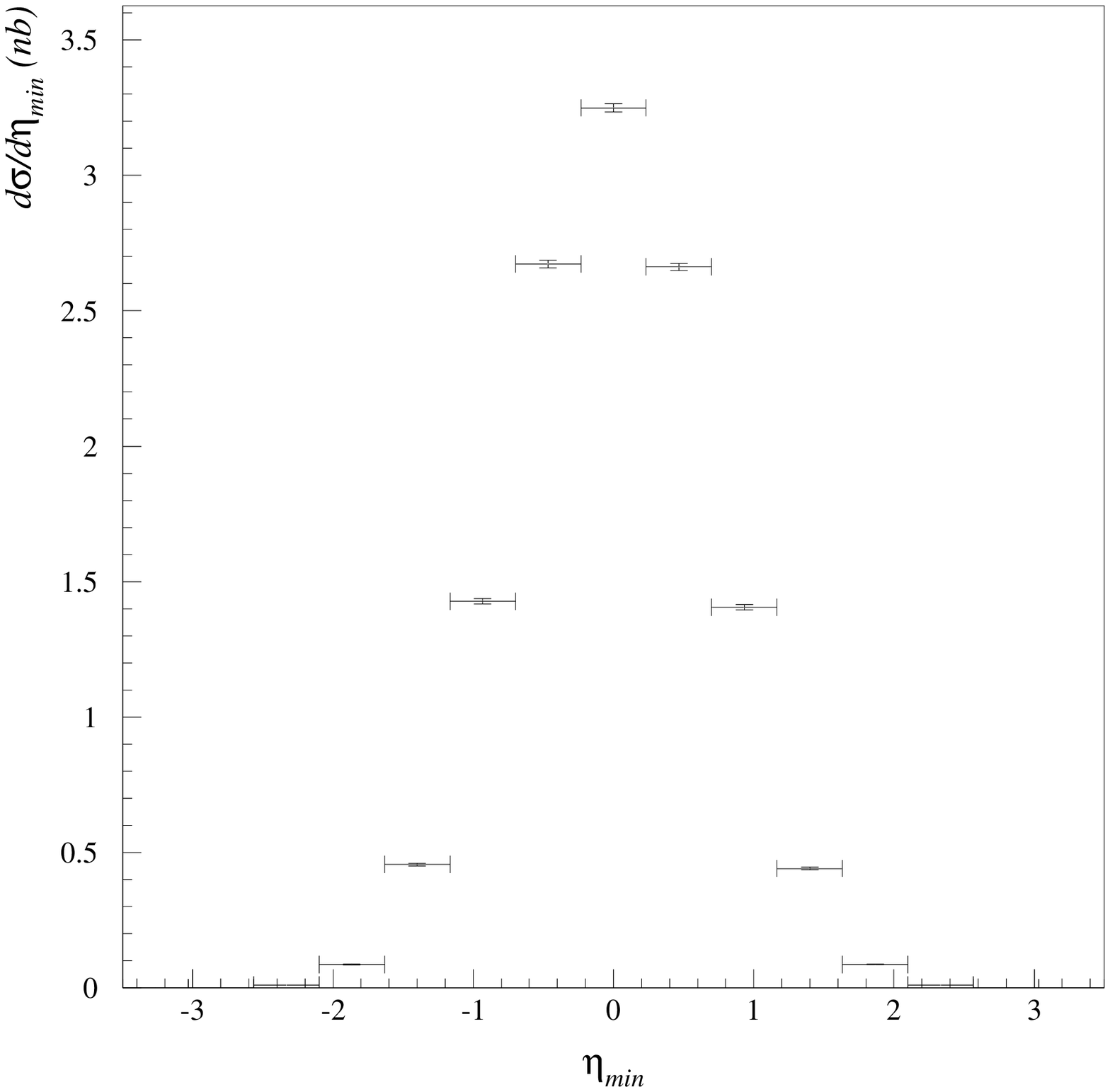,width=55mm,height=35mm}
\end{center}
\caption{\it Rapidity distribution of the most (left panel)
and the least (right panel) energetic gluon in the $gg\rightarrow 4g$ process
with MC summation over colour.}
\label{eta_6g}
\end{figure}

In this section, several numerical results for multi-parton production at
the LHC are presented. The main aim is to show that the MC  summation
over colour,  which speeds up the calculation enormously,  gives
results with precision comparable to the one based on explicit
summation.  The centre of mass energy was chosen to be $\sqrt{s}=14$
$\textnormal{TeV}$.  In order to remain far from collinear and  soft
singularities and to simulate as much as possible the experimentally
relevant phase-space regions, we have chosen the following cuts:
\begin{equation}
\label{cuts}
p_{T_{i}} > 60 ~\textnormal{GeV}, ~~~~ |y_{i}|<2.5, ~~~~ \Delta R_{ij}
> 1.0
\end{equation}
for each pair of outgoing partons $i$ and $j$. 
All results are obtained with a fixed strong coupling
constant ($\alpha_s$=0.13).  For the parton structure functions, we
used the {\tt CTEQ6 PDF}'s parametrisation
\cite{Pumplin:2002vw,Stump:2003yu}.  For the phase space generation we
used  the algorithm described in Ref. \cite{Papadopoulos:2005ky},   whereas
in all cases results were cross checked with {\tt PHEGAS}
\cite{Papadopoulos:2000tt},  {\tt HAAG}\cite{Hameren:2002tc} and {\tt
RAMBO} \cite{Kleiss:1985gy}.  In  Tab.\ref{tab1} the results for
the total cross section for processes with gluons are presented.  All
cross sections are in agreement within errors.
\begin{table}[h!]
\tbl{Results for the total cross section
for processes with gluons only.
$\sigma_{\textnormal{{\tiny EXACT}}}$ corresponds to summation over
all possible colour configurations, while
$\sigma_{\textnormal{{\tiny MC}}}$ corresponds to
 MC summation.}
{\normalsize
\begin{tabular}{@{}crr@{}}
\hline & &\\ 
$\textnormal{Process}$ & $\sigma_{\textnormal{\tiny EXACT}}$
$\pm$ $\varepsilon$ $\textnormal{(nb)}$& 
$\sigma_{\textnormal{\tiny MC}}$
$\pm$ $\varepsilon$ $\textnormal{(nb)}$ \\  
 & & \\  \hline & & \\  
$gg \rightarrow 2g$ & (0.46572 $\pm$ 0.00258)$\times 10^{4}$&
(0.46849 $\pm$ 0.00308)$\times 10^{4}$ \\
$gg \rightarrow 3g$ & (0.15040 $\pm$ 0.00159)$\times 10^{3}$
&(0.15127 $\pm$ 0.00110)$\times 10^{3}$\\
$gg \rightarrow 4g$  & (0.11873 $\pm$ 0.00224)$\times 10^{2}$
&(0.12116 $\pm$ 0.00134)$\times 10^{2}$\\
$gg \rightarrow 5g$ & (0.10082 $\pm$ 0.00198)$\times 10^{1}$
&(0.09719 $\pm$ 0.00142)$\times 10^{1}$\\
$gg \rightarrow 6g$ &(0.74717 $\pm$ 0.01490)$\times 10^{-1}$
&(0.76652 $\pm$ 0.01862)$\times 10^{-1}$ \\ & & \\
\hline
\end{tabular}
\label{tab1}}
\end{table}
Rapidity and transverse momentum distributions  of the most and the
least energetic  parton for $gg\rightarrow 4g$  processes are shown in
Fig.\ref{pt_6g} and Fig.\ref{eta_6g}.  They clearly demonstrate that
MC summation over colour  performs very well not only at the level of
total rates but also at the level of differential distributions.

\section{Summary}
A status report on an efficient tool for  automatic computation of
helicity amplitudes and cross sections for multi-jet final states in
the SM with the LHA event record interface for parton shower and
hadronisation  to {\tt PYTHIA} has been shortly presented.

\section*{Acknowledgments}
This work is partly supported by the EU grant MTKD-CT-2004-510126 in
partnership with the CERN Physics Department and by the Polish
Ministry of Scientific Research and Information Technology grant No
620/E-77/6.PR UE/DIE 188/2005-2008. The Greece-Poland bilateral
agreement {\it ``Advanced computer techniques for theoretical
calculations and development of simulation programs for high energy
physics experiments'' } is also acknowledged.


\providecommand{\href}[2]{#2}

\end{document}